 \documentclass[final,3p,times,twocolumn]{elsarticle}

\usepackage{amssymb}

\usepackage{color}
\usepackage{graphicx}
\usepackage{graphicx,epstopdf}
\usepackage{dcolumn}
\usepackage{bm}
\usepackage{mathrsfs}
\usepackage{soul}
\usepackage{mathtools}
\usepackage{blkarray, bigstrut}
\usepackage{hyperref}

\usepackage{mathtools}

\journal{Journal of Theoretical Biology}

\begin{document}

\begin{frontmatter}



\title{Eco-evolutionary dynamics of cooperation in the presence of policing}

\author{Sayantan Nag Chowdhury$^{1,*}$, Srilena Kundu$^{1, *}$, Jeet Banerjee$^2$, Matja{\v z} Perc$^{3,4,5}$ and Dibakar Ghosh$^1$}
\address{$^1$Physics and Applied Mathematics Unit, Indian Statistical Institute, 203 B. T. Road, Kolkata-700108, India\\
	$^2$BYJU'S, Think \& Learn Pvt. Ltd., IBC Knowledge Park, 4/1 Bannerghatta Main Road, Bangalore - 560029, India\\
	$^3$Faculty of Natural Sciences and Mathematics, University of Maribor, Koro{\v s}ka cesta 160, 2000 Maribor, Slovenia\\
	$^4$Department of Medical Research, China Medical University Hospital, China Medical University, Taichung, Taiwan\\
	$^5$Complexity Science Hub Vienna, Josefst{\"a}dterstra{\ss}e 39, 1080 Vienna, Austria\\
    $^*$ Both the authors have contributed equally to the work}
\ead{dibakar@isical.ac.in (D. Ghosh), matjaz.perc@gmail.com (M. Perc), jeet.banerjee@byjus.com, jeetban89@gmail.com (J. Banerjee)}

\title{}


\author{}

\address{}

\begin{abstract}
Ecology and evolution are inherently linked, and studying a mathematical model that considers both holds promise of insightful discoveries related to the dynamics of cooperation. In the present article, we use the prisoner's dilemma (PD) game as a basis for long-term apprehension of the essential social dilemma related to cooperation among unrelated individuals. We upgrade the contemporary PD game with an inclusion of evolution-induced act of punishment as a third competing strategy in addition to the traditional cooperators and defectors. In a population structure, the abundance of ecologically-viable free space often regulates the reproductive opportunities of the constituents. Hence, additionally, we consider the availability of free space as an ecological footprint, thus arriving at a simple eco-evolutionary model, which displays fascinating complex dynamics. As possible outcomes, we report the individual dominance of cooperators and defectors as well as a plethora of mixed states, where different strategies coexist followed by maintaining the diversity in a socio-ecological framework. These states can either be steady or oscillating, whereby oscillations are sustained by cyclic dominance among different combinations of cooperators, defectors, and punishers. We also observe a novel route to cyclic dominance where cooperators, punishers, and defectors enter a coexistence via an inverse Hopf bifurcation that is followed by an inverse period doubling route.
\end{abstract}

\begin{graphicalabstract}
\end{graphicalabstract}

\begin{highlights}

\item A theoretical model is designed to bridge the gap between ecology and evolution.	

\item Prisoner's dilemma game is upgraded  using the ecological signature of free space.

\item Interplay of punishment, altruist free space, and mortality rate is revealed.

\item The eco-evolutionary model leads to fascinatingly different dynamics.

\item Periodic attractor reveals cyclic dominance among different subpopulations.

\end{highlights}


\end{frontmatter}



\section{Introduction}

\par Competition, within and between species for their existence and survivability \cite{murray2007mathematical,banerjee2019supercritical,kundu2017survivability}, is one of the fundamental attributes under the realm of Darwinian theory of evolution \cite{darwin1909origin}. The possible persistence and rapid emergence of non-cooperative strategy \cite{smith1997major} challenge the cooperative contribution in the presence of defectors and it leads to the ``tragedy of the commons" \cite{Hardin1243} as only the fittest are most likely to overcome the fierce struggles of life. This mechanism of survival of the fittest generates an act of selfishness among the individuals \cite{sigmund2010calculus}, which hinders the evolution and maintenance of cooperation. Surprisingly, contradictory to the famous Darwinian evolutionary theory, cooperation among self-interested individuals is found in diverse circumstances ranging from microbial populations to social systems \cite{axelrod1984evolution,perc2017statistical,fotouhi2019evolution,tanimoto2007emergence,axelrod1981evolution,tanimoto2007does,fotouhi2018conjoining}. Cooperation is often observed in the community of birds in the form of taking care of other's offsprings \cite{skutch1961helpers}. Large-scale cooperative behavior is very common even in simple organisms like bee and ant \cite{wilson1971insect,wang2008trade}, which can form captivating things, such as shaft systems to ventilate their nests. A series of publications \cite{nowak2006five,perc2010coevolutionary,javarone2015emergence,szabo2007evolutionary,wakano2009spatial,nag2020cooperation,perc2013evolutionary,pennisi2005did,wang2015evolutionary,santos2005scale,gomez2007dynamical,fu2008reputation,wakano2011pattern,wu2011moving,zhang2013tale} have been produced by the scientific communities across various disciplines for understanding the mechanisms behind the initiation, emergence and promotion of cooperation.

\par Evolutionary game theory \cite{weibull1995evolutionary,nowak2006evolutionary,smith1982evolution,hofbauer1998evolutionary}, one of the powerful competent  theoretical frameworks for analyzing the long-standing puzzle on the evolution of cooperation in public goods games, applies the mathematical theory of games in the contexts of biological and social systems. Two-player games have become the general relevant metaphor, which help to shed light on the paradigm for studying the emergence of widespread coordination under the paradox between collective and individual rationality. In these simultaneous pair-wise interaction games, the outcome of an individual depends solely on the chosen strategies of the opponent and the player itself. 
 One of the general examples of such $2 \times 2$ games is prisoner's dilemma (PD) game \cite{axelrod1981evolution}. The pairwise mutual interaction between the players generates a strategy space containing four possible payoff values. Two players may simultaneously decide either to cooperate or to defect without any prior knowledge of other player's choices. A defector exploiting a cooperator gets a temptation amount $T$ and the exploited cooperator receives the sucker's payoff $S$. They will both receive the reward $R$ and
punishment $P$ for mutual cooperation and mutual defection, respectively. Generally, the payoffs in the PD game satisfy the inequalities $T > R > P > S$ and $2R > T+S$ \cite{szabo2007evolutionary}. Clearly, these inequalities suggest that players need to defect, irrespective of opponent's strategy, for guaranteed highest income in terms of their own payoff. Naturally, if both defect, they will get $P$, that is comparatively lower than $R$, which they would have obtained when they both cooperate. This scenario leads to the emergent dilemma, and as a result of that, widespread defection is the natural unfortunate outcome failing to sustain cooperation in the classical well-mixed PD game. The PD game is capable of capturing the notions of several other social dilemma games \cite{liebrand1983classification,poundstone1992prisoner}. Recent progress in evolutionary game theory identified various mechanisms that support the evolution of cooperation \cite{nowak2006five}.

\par Punishment (Policing) \cite{brandt2006punishing,cong2017individual,wang2013impact,banerjee2019delayed,yang2015peer,helbing2010defector,perc2012sustainable,dreber2008winners,yang2015role} is one of the effective mechanisms, which helps to achieve global and individual optima (evolutionary stable equilibrium) of cooperation under suitable circumstances. Besides two distinct strategies, cooperation and defection, an additional strategy punishment is introduced, which challenges the free-riding behavior of the defectors and entertains the maintenance of cooperation \cite{fehr2002altruistic}. Punishers are also cooperators, but they differ from traditional cooperators (``second-order free riders" \cite{ozono2016solving,szolnoki2017second}) by imposing a cost in terms of payoff towards restricting the unimpeded exploitation of cooperative goods by the free-riders. The evidence of punishment is ubiquitous in not only human society but also unicellular bacterial community \cite{banerjee2019delayed}. The cooperative producers secrete toxins (i.e., hydrogen cyanide) to mitigate the unrestricted usage of public goods, such as elastase, by toxin-sensitive non-productive defective mutants. The act of policing leaves two distinct alternatives to the defectors on how to proceed. Defectors may still continue to defect with the hope that natural selection ultimately favors defection as compared to cooperation, or they may decide to cooperate leading to a situation which is the best for the group. However, a recent study \cite{jiang2013if} reveals that mild punishment may be more effective in improving selfless cooperative behaviors. It should be noted that punishment is not a mechanism for the evolution of cooperation \cite{nowak2006five}. In fact, most of these earlier investigations \cite{helbing2010evolutionary,fowler2005altruistic,szolnoki2011phase} are confined to public goods game, and little is known regarding the possible evolutionary impact of punishment on the dynamics of PD game. Various aspects of punishment are already scrutinized by means of different experiments \cite{egas2008economics,sasaki2007probabilistic} and mathematical models \cite{henrich2001people,bowles2004evolution,ohtsuki2009indirect,hauert2002replicator,brandt2006good}.

\par {Ecologists and evolutionary biologists typically assume that evolutionary processes are much slower than ecological processes \mbox{\cite{slobodkin1980growth}}. However, recent studies show that ecological changes and species evolution can occur on the same time scale, i.e., ecological and evolutionary dynamics are intertwined \mbox{\cite{hendry2020eco}}. Ecological changes can significantly impact evolutionary dynamics, and the resulting evolutionary changes can feedback on the ecological dynamics \mbox{\cite{colombo2019spatial}}. We are at a stage, where the consideration of intimate interlinking between ecology and evolution is a necessary step for the understanding of the processes that regulate biodiversity \mbox{\cite{pelletier2009eco}}.} 

\par In the present article, we explore the interplay between the punishment and the virtual ‘optional discriminatory’ altruistic behavior of the free space from a somewhat different perspective. Altruism \cite{de2010freedom,bell2008selection} refers to the selflessness of individual, who increases the fitness of another individual, either directly or indirectly, without the expectation of reciprocity for that action. Evolutionary social behaviors are omnipresent in nature, and the impact of ecological free space may be a crucial factor in the context of eco-evolutionary dynamics \cite{pelletier2009eco,wang2020eco,fussmann2007eco,wang2020steering}. We consider free space as an ecological variable, which can be occupied by an offspring of any subpopulation of cooperators, defectors and punishers. So, by losing its own identity, free space provides benefit to all other individuals and most importantly, it does not take any form of advantages from others. The role of free space \cite{nag2020cooperation,armano2017beneficial,helbing2009outbreak,aktipis2004know,smaldino2012movement,meloni2009effects,vainstein2007does} receives a great deal of attention under the framework of evolutionary game theory. But, the interdependency between altruist free space and the social punishment has been largely unexplored in the existing literature. We adopt a modeling approach by describing the temporal evolution of the densities of the different subpopulations. Our finding suggests that the selfless one-sided contribution of altruist free space leads to various emergent attractors \cite{perko2013differential,nag2020hidden}.

\par We add another layer of complexity by introducing a natural per capita mortality rate \cite{finch2010evolution,burger2012human,wachter1997evolutionary}. Many studies have extensively demonstrated the impact of several factors like social, economic, and health implications on the reductions in mortality \cite{brayne2007elephant,oeppen2019life}. An elementary discussion, concerning the implication of mortality change for evolutionary theories of PD game, is yet to gain its well-deserved attention. {We formulate a general mathematical model in the presence of evolutionary social behavior, punishment, to address the combined effect of altruistic free space and mortality change on the evolution of population. We also hope that our research exhibits a better understanding of eco-evolutionary dynamics in social dilemmas.} Consideration of all these aspects unveils the coexistence of three competing strategies under favorable conditions, and prompts the emergence of cyclic dominance \cite{szolnoki2014cyclic}, where the population system displays a periodic attractor. {Emergence of such periodic attractor through Hopf bifurcation has been studied earlier in eco-evolutionary models} \cite{gokhale2016eco,cortez2016magnitude}. We, hereafter, proceed by investigating the evolutionary dynamics among cooperators, defectors, and punishers in an infinite population and provide a rigorous stability analysis of the system. The presented theoretical investigations are well agreed with our numerical studies. The system experiences two clearly separated time scales consisting of fast jumps followed by a slow manifold \cite{nag2020hidden,arnold1998random}. The stability properties of the proposed mathematical model are further numerically analyzed by bifurcation theory and Lyapunov exponents of the system.

\section{Mathematical Model: Eco-evolutionary dynamics}

\par We consider our model based on the repeated prisoner's dilemma game. {The basic game consists of two possible behaviors, cooperation, $\mathbf{C}$, and defection, $\mathbf{D}$; but, we include punishment to extend the set of strategies to three distinct behaviors, $\mathbf{C}$, $\mathbf{D}$, and punishment $\mathbf{P}$.} Instead of the traditional PD game, the weak version of the prisoner's dilemma game \cite{nowak1992evolutionary} is contemplated, where the rank of the payoffs between $\mathbf{C}$ and $\mathbf{D}$ are characterized by $T > R > P \geq S$. Without any loss of generality, we set $R=1$, $S=0$, $T=\beta$ with $\beta > 1$, and $P=0$, which helps to preserve the dilemma of the weakly PD game. The punishers ($\mathbf{P}$s) impose a fine on defectors at a personal cost. At the time of interaction with a punisher, defectors have to bear a punishment fine $\delta$, and punishers also endure the same cost of policing, $\delta$. Thus, $\delta > 0$. {Since punishers ($\mathbf{P}s$) are cooperative in nature, a punisher ($\mathbf{P}$) and a cooperator ($\mathbf{C}$) both receive the reward $R=1$ due to the mutual interaction between them.}

\par In order to combine the game dynamics with the population dynamics, we consider free space as an ecological variable, which interacts with all other subpopulations $\mathbf{C}$, $\mathbf{P}$ and $\mathbf{D}$. {Free space does not take any advantage from others, but any subpopulation can use free space for their replication, i.e., free space is providing benefit to all $\mathbf{C}$, $\mathbf{P}$ and $\mathbf{D}$.} Moreover, when free space is occupied by an offspring of $\mathbf{C}$, $\mathbf{P}$ or $\mathbf{D}$, then it loses its identity.  Therefore, we can assume that the free space is such a behavior, which selflessly increases the fitness of other subpopulations and eliminates its own identity. It needs to be mentioned that free space can be surrounded by
cooperators or cooperative-punishers or defectors. Hence, our initial assumption that the free space is also interacting with other subpopulations, such as $\mathbf{C}$, $\mathbf{P}$ and $\mathbf{D}$, allows us to depict the selfless act of free space as virtual ‘optional discriminatory’ altruistic behavior. The altruistic act of free space $\mathbf{F}$ allows it to contribute positive payoff $\sigma_{1}$, $\sigma_{2}$ and $\sigma_{3}$ to $\mathbf{C}$, $\mathbf{P}$ and $\mathbf{D}$, respectively. Therefore, $\sigma_{i} > 0$ for $i=1,2$, and $3$. The payoff matrix is therefore represented by

\[
\begin{blockarray}{ccccc}
\hspace{0.1 cm} & \mathbf{C} & \mathbf{P} & \mathbf{D} & \mathbf{F} \\
\begin{block}{c(cccc)}
 \mathbf{C} \hspace{0.1 cm} & 1 & 1 & 0 & \sigma_{1}  \\
 \mathbf{P} \hspace{0.1 cm} & 1 & 1 & -\delta & \sigma_{2} \\
 \mathbf{D} \hspace{0.1 cm} & \beta & \beta-\delta & 0 & \sigma_{3} \\
\mathbf{F} \hspace{0.1 cm}  & 0 & 0 & 0 & 0 \\
\end{block}
\end{blockarray}
\]

in which the entries portray the payoff accumulated by the players in the left.

\par {Let $x$, $y$, $z$ and $w$ be the fraction of $\mathbf{C}$, $\mathbf{P}$, $\mathbf{D}$ and $\mathbf{F}$, respectively.} {It is assumed that the community is only comprised of $\mathbf{C}$, $\mathbf{P}$, $\mathbf{D}$ and $\mathbf{F}$, therefore, $x + y + z + w = 1$.} As $w$ is a virtual ‘optional discriminatory’ altruist, the normalized population density becomes $x + y + z$. 
The overall population density, $x + y + z$, can grow  from $0$ to an absolute maximum $1$. {If $x+y+z = 0$, i.e., $w = 1$, then only free space will be available and population extinction will occur. If $x+y +z = 1$, i.e.,
$w = 0$, then there will be no free space and the maximum normalized population density exists.} Therefore, $0 \leq x + y + z \leq 1$, i.e., we consider the varying normalized population density. Using the payoff matrix, the average fitness of {each subpopulation can be calculated.}

\par The average fitness of $\mathbf{C}$ is given by
\begin{equation} \label{1}
\begin{array}{lcl}
f_{C} =x + y + \sigma_{1} w=(1 - \sigma_{1} )x + (1 - \sigma_{1} )y - \sigma_{1} z + \sigma_1,
\end{array}
\end{equation}
where the relation $w = 1 - x - y - z$ is used to eliminate the dependent variable $w$, i.e., the fraction of available freespace, which explicity depends on the abundances of constituent subpopulations, \textbf{C}, \textbf{P}, and \textbf{D}.



Similarly, the respective average fitness of $\mathbf{P}$ and $\mathbf{D}$ are
\begin{equation} \label{2}
\begin{array}{lcl}
 f_{P}=(1 - \sigma_{2})x + (1 - \sigma_{2} )y - (\delta + \sigma_{2} )z + \sigma_{2},
\end{array}
\end{equation}
and
\begin{equation} \label{3}
\begin{array}{lcl}
f_{D}=(\beta - \sigma_3 )x + (\beta - \delta - \sigma_3 )y - \sigma_3 z + \sigma_3 .
\end{array}
\end{equation}

\par Thus, the fractions $x$, $y$ and $z$ determine the average payoffs $f_{C}$, $f_{P}$ and $f_{D}$ of cooperators, punishers and defectors, respectively, at any given point of time. As we have assumed that free space is not taking any advantage from others (notion of altruistic behavior), the average fitness of $\mathbf{F}$ can be denoted by

\begin{equation} \label{4}
\begin{array}{lcl}
f_{F}=0.
\end{array}
\end{equation}

\par The average payoff of the entire population is

\begin{equation} \label{5}
\begin{array}{lcl}
\hspace{0.8cm} \bar{f}=\dfrac{xf_{C}+yf_{P}+zf_{D}}{x+y+z}= \dfrac{xf_{C}+yf_{P}+zf_{D}}{1-w}.
\end{array}
\end{equation}

\par To determine the dynamics of $x$, $y$ and $z$, we assume that all individuals die at an equal and common rate $\xi$ {and, to reduce the complexity of the system, we assume that the reproduction rate is fully controlled by $f_{C}$ , $f_{P}$ and $f_{D}$.} 
 Thus, the eco-evolutionary dynamics of $\mathbf{C}$, $\mathbf{P}$, $\mathbf{D}$ and $\mathbf{F}$ can be expressed as

\begin{equation} \label{6}
\begin{array}{lcl}
\dot{x} =x(f_{C}-\xi),\\
\dot{y} =y(f_{P}-\xi),\\
\dot{z} =z(f_{D}-\xi),\\
\dot{w}=-\dot{x}-\dot{y}-\dot{z}.
\end{array}
\end{equation}

\begin{figure*}[ht]
	\centerline{\includegraphics[scale=0.6]{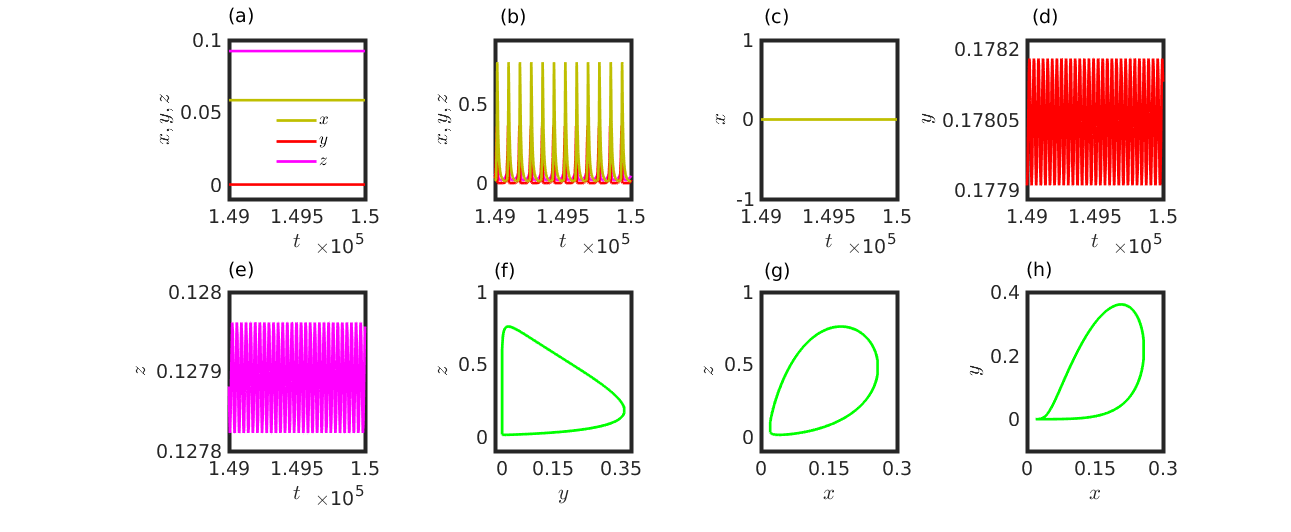}}
	
	 \caption{{\bf Example of Eco-evolutionary dynamics of the system \eqref{7} for different initial conditions}: The system {\mbox{\eqref{7}}} possesses at most three types of multistability for our chosen parameter values of $\sigma_{1}=0.52$, $\sigma_{2}=0.72$, $\sigma_{3}=0.41$, $\xi=0.5$, $\delta=1.39$, and $\beta=2.6$. The initial condition $(x_0,y_0,z_0)$ is set at (a) $(0.7,0,0.2)$, (b) $(0.3,0.3,0.3)$, (c)-(e) $(0,0.2,0.7)$, and (f)-(h) $(0.4,0.2,0.3)$, respectively. (a) The punisher-free $(y=0)$ stationary state $E_{5}$ {even in the presence of moderate punishment ($\delta = 1.39$). The dominance of defectors over the cooperators is observed, as $z > x$, even though the initial fraction of cooperators, $x_0$, is higher than the initial fraction of defectors, $z_0$.} (b), (f)-(h) Periodic oscillation of the frequencies of $\mathbf{C}$, $\mathbf{P}$ and $\mathbf{D}$ {for initial conditions with non-zero components. Even when the punishers are given less favorable platform as $x_0 > z_0 > y_0$ ((f)-(h)) and the temptation to defect is high ($\beta = 2.6$), cooperation is still effectively sustained under adverse conditions while strategy abundances keep oscillating which prompts the emergence of cyclic dominance.} (c)-(e) Extinction of cooperation ($x=0$) with small amplitude oscillation of $y$ and $z$. {The relation between punishers and defectors in the absence of cooperators provides an emergent oscillatory dynamics, where interestingly punishers dominate the defectors, as $y > z$, in spite of the given initial preferences towards defectors.} For further simulation details, please see the text. }\label{fig1}
\end{figure*}

 Clearly, the changes in frequencies of all subpopulations over time, governed by the Eq.\ \eqref{6}, can be thought of as an extension of replicator dynamics \cite{hofbauer1998evolutionary}, as by setting $\xi=\bar{f}$ (where $\bar{f}$, the mean fitness, is given by the Eq.\ \eqref{5}), one can easily recover the traditional replicator system. {It needs to be mentioned that in general free space gives a positive feedback to the growth-induced reproduction of a population; hence, in the proposed approach, we consider the per-capita growth rate of each of the subpopulations \textbf{C}, \textbf{P}, and \textbf{D} is dependent on the availability of accessible free space.} It is clear from the model formation that we have already considered the fraction of free space $w$ and respective benefits $\sigma_{1}$, $\sigma_{2}$, $\sigma_{3}$, in the fitnesses of subpopulations, which signify their reproduction rate. That is why, we have omitted the redundant multiplication of $w$ with $f_{C}$, $f_{P}$ and $f_{D}$, which is often observed in the previous studies \cite{gokhale2016eco,hauert2006evolutionary}.

\par After substituting $f_{C}$ (Eq.\ \eqref{1}), $f_{P}$ (Eq.\ \eqref{2}) and $f_{D}$ (Eq.\ \eqref{3}) in dynamics \eqref{6}, we obtain the following eco-evolutionary dynamics

\begin{equation} \label{7}
\begin{array}{lcl}
\dot{x} =x\left[(1-\sigma_1)x+(1-\sigma_1)y-\sigma_1 z +(\sigma_1-\xi)\right],\\
\dot{y} =y\left[(1-\sigma_2)x+(1-\sigma_2)y-(\sigma_2+\delta)z +(\sigma_2-\xi)\right],\\
\dot{z} =z\left[(\beta-\sigma_3)x+(\beta-\sigma_3-\delta)y-\sigma_3 z +(\sigma_3-\xi)\right],
\end{array}
\end{equation}
where, $\sigma_{1}$, $\sigma_{2}$, $\sigma_{3}$, $\delta$, $\xi > 0$ and $\beta > 1$.

\section{Results}

\subsection{Model calibration and analysis}

\par  To explore the dynamics of the system \eqref{7}, rigorous numerical experiments have been performed for a wide range of six parameters $\sigma_{1}$, $\sigma_{2}$, $\sigma_{3}$, $\xi$, $\beta$ and $\delta$. The fifth-order Runge-Kutta-Fehlberg method is used to integrate the system \eqref{7} with a fixed time step $0.01$. To avoid computational error due to sensitive initial data, we observe the evolution of trajectories after sufficient initial transient of $1.3 \times 10^7$ iteration steps. Detailed theoretical analysis is shown in \ref{Positivity and boundedness of solutions}, ensuring the positive invariance and uniqueness of the solutions of the model \eqref{7}. The analytical conditions for existence and stability of various equilibria of the system \eqref{7} are also analyzed in \ref{Existence and stability analysis of the stationary state} using standard methods of linear stability analysis.

\subsection{Temporal behavior of the densities of three subpopulations}

\par The initial fraction of free space $\mathbf{F}$ is kept fixed at $w_{0}=0.1$ \footnote{{The initial fraction of free space, $w_{0}$, does not qualitatively affect the numerical findings obtained.} $w_{0}$ can be varied within the closed interval $[0,1]$ obeying the relation $x_0+y_0+z_0+w_0=1$. If $w_{0}=1$, then the initial fraction of subpopulations is reduced to a singleton choice $(x_0,y_0,z_0)=(0,0,0)$ and hence all the species will die out giving rise to the stationary point $E_{0}$. As $w_{0} \to 0+$, then the region of initial basin consisting $x_{0}$, $y_{0}$ and $z_{0}$ is expanded, and $x_{0}+y_{0}+z_{0} \to 1-$.  }. Thus, the initial individual densities of different subpopulations can be varied within the interval $[0,0.9]$ maintaining the relation $x_{0}+y_{0}+z_{0}=0.9$. To investigate the evolutionary dynamics, without loss of any generality, we fix the values of all the parameters at $\sigma_{1}=0.52$, $\sigma_{2}=0.72$, $\sigma_{3}=0.41$, $\xi=0.5$, $\delta=1.39$, and $\beta=2.6$. Interestingly, we observe different emergent dynamical behavior of the attractor solely based on the choices of initial conditions. {A glimpse of this scenario is portrayed in} Fig.\ \ref{fig1}. For example, the initial choice of $(x_{0}, y_{0}, z_{0})=(0.7,0,0.2)$ leads to the extinction of punishers $\mathbf{P}$ {even in the presence of moderate punishment ($\delta=1.39$).} The temporal evolution of the trajectories (See Fig.\ \ref{fig1} (a)) {depicts that the system} \eqref{7} converges to {the punisher-free stationary point ($E_{5}$). In the absence of punishers, the dominance of defectors over the cooperators ($z>x$) is well expressed at the steady-state of coexistence} {even though the initial fraction of cooperators, $x_0$, is higher than the initial fraction of defectors, $z_0$.} {For the same set of parameter values, the initial fraction $(0.7,0.2,0)$ of all subpopulation gives rise to unbounded solution of the system} \mbox{\eqref{7}} {(Figure not shown here). Due to non-uniformity in the altruistic reproductive benefit of free space to the cooperators and punishers ($\sigma_{1} \ne \sigma_{2}$)}, {the defector-free stationary steady state ($E_{4}$) can not be obtained in this case (for a detailed discussion, please see \mbox{\ref{Existence and stability analysis of the stationary state}}).}

 \par {For interior initial conditions, i.e., initial conditions with non-zero components, the system exhibits a periodic attractor.} For instance, equal probabilities of initial fraction $(x_0,y_0,z_0)=(0.3,0.3,0.3)$ generate such periodic trajectories, shown in Fig.\ \ref{fig1} (b). Similar state-space diagram, projected onto the two-dimensional space, is contemplated for the initial condition $(x_0,y_0,z_0)=(0.4,0.2,0.3)$ in the Figs.\ \ref{fig1} (f)-(h). {Although the punishers are given less favorable platform as $x_0 > z_0 > y_0$, the post-transient eco-evolutionary dynamics depict these subpopulations indeed cyclically dominate one another in the irregular mixing pattern of trajectories. Note that, even though the temptation to defect is high ($\beta=2.6$), cooperation is still effectively sustained under adverse conditions while strategy abundances keep oscillating.} This prompts the emergence of cyclic dominance, whereby defectors dominate punishers who dominate cooperators who in turn, dominate defectors. Besides cyclic dominance, two clearly disjoint time scales consisting of fast jumps followed by a slow manifold are evidenced in these Figs.\ \ref{fig1} (b) and \ref{fig1}(f)-(h). {The trajectories slowly approach the origin when in its neighborhood, but, when in close vicinity of the origin, the trajectories are leaving comparatively fast generating a distinct slow-fast time scale.}

\par Focusing on a different scenario, we choose another initial condition $(x_0,y_0,z_0)=(0.0,0.2,0.7)$. Clearly, here the initial fraction of defectors are sufficiently high, giving the defectors initial advantage. As $x_0=0$, the density of cooperators will remain zero ($x=0$) as shown in Fig.\ \ref{fig1} (c). The relation between punishers and defectors in the absence of cooperators provides an emergent oscillatory coexistence between punishers and defectors, where interestingly punishers dominate the defectors, as $y>z$, in spite of the given initial preferences towards defectors. The trade-off between temptation ($\beta=2.6$) and punishment ($\delta=1.39$) might be a cause towards the steady-state domination of punishers over defectors in the absence of unrestricted defective exploitation of cooperation. A small amplitude oscillations of $y$ and $z$ are visible through Figs.\ \ref{fig1} (d) and (e).

\begin{figure*}[ht]
	\centerline{\includegraphics[scale=0.55]{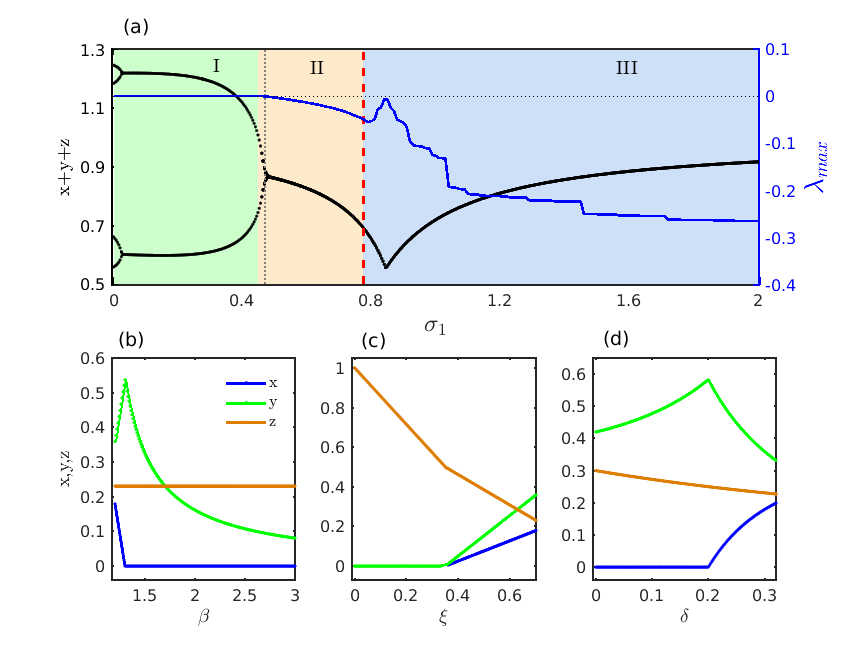}}
	\caption{{\bf (a) The largest Lyapunov exponent $\lambda_{max}$ and bifurcation diagram}: {The frequency of sum of all three subpopulations is depicted as a function of $\sigma_1$ through the black bifurcation curve that reflects an inverse Hopf route corresponding to the destruction of the periodic orbit. Depending on the values of the normalized population density $x + y + z$, the entire bifurcation diagram is partitioned into three distinct sub-regions.} Region I: oscillatory coexistence of $x, y, z <1$ but $x+y+z >1$ and $y>z>x$. Region II: Oscillatory + stable coexistence of $x, y, z$ and $y>z>x$ until $\sigma_1 = 0.775$. Red dashed curve: $x = y = z$ at $\sigma_1 = 0.775$ (stable coexistence). Region III: $x>z>y$, stable coexistence of $x,y,z$ till $\sigma_1 = 0.844$ (bifurcation point), after that coexistence of $x,z$ ($y=0$, $x>z>y$). {Higher values of $\sigma_1$ help to sustain cooperation depending on values of the other parameters. As soon as $\sigma_1$ crosses beyond $\sigma_2$ and $\sigma_3$, i.e., $\sigma_1 > \sigma_2$ and $\sigma_1 > \sigma_3$; the free space is providing greater benefits to the cooperators compared to the punishers and defectors. To further validate the appearance of oscillatory or stable coexistence of the populations, the largest Lyapunov exponent $\lambda_{max}$ is plotted in blue curve by varying $\sigma_1$. The sign of $\lambda_{max}$ changes from $0$ to -ve value, assuring the transformation of the attractor from periodic nature to stationary state.} {\bf (b)-(d) Individual effect of the parameters $\beta$, $\xi$ and $\delta$ on $x,y,z$}: Fraction of cooperators, punishers, and defectors as a function of temptation parameter $\beta$, per capita mortality rate $\xi$ and policing parameter $\delta$ for fixed $\sigma_1 = 0.7$. {(b) Higher values of temptation parameter $\beta$ always helps to provide additional benefits to the defectors and destroys the act of cooperation, (c) increment of mortality rate $\xi$
surprisingly encourages the coexistence of all subpopulations, (d) policing $\delta$ of appropriate strength can fight the free-rider's defection and promote cooperation in the long run.} All the results are carried out with fixed initial fraction of population $(0.3,0.3,0.3)$ and the parameter values are fixed at: $\sigma_2 = 1.0$, $\sigma_3 = 0.7$, $\beta = 1.2$, $\delta = 0.3$, and $\xi = 0.7$.}\label{fig2}
\end{figure*}

\subsection{Interplay of different parameters}

\par We observe different dynamical characteristics in our model depending on initial fraction of species and different parameters. In order to further understand the role of parameters behind the results as presented in Fig.\ \ref{fig1}, we analyze the frequency of sum of all three subpopulations as a function of free space induced reproductive benefit to the cooperators, $\sigma_1$, with fixed initial condition $(0.3,0.3,0.3)$. It is highly anticipated that higher values of $\sigma_1$ help to sustain cooperation; however, it additionally depends on the values of other parameters. As soon as $\sigma_{1}$ exceeds $\sigma_2$ and $\sigma_3$, i.e., $\sigma_{1} > \sigma_{2}$ and $\sigma_{1} > \sigma_{3}$; the free space is providing greater benefits to the cooperators compared to the punishers and defectors. This feature is well manifested in Fig.\ \ref{fig2} (a) {keeping the parameter values fixed at
$\sigma_2 = 1.0$, $\sigma_3 = 0.7$, $\beta = 1.2$, $\delta = 0.3$, and $\xi = 0.7$.} The black bifurcation curve reflects an inverse Hopf route corresponding to the destruction of the periodic orbit. Depending on the values of the normalized population density $x+y+z$, we are able to partition the entire bifurcation diagram into three distinct sub-regions. Region I of Fig.\ \ref{fig2} (a) contemplates the oscillatory coexistence of all three subpopulations. The temporal dynamics (not shown here) at a particular time snapshot suggests that frequencies of three strategies oscillate with $y>z>x$ and thus, punishment strategy can be dominant. In spite of achieving such delightful persistence of all subpopulations, we have to ignore this regime, as $x+y+z>1$ signifying overcrowded population within this regime. Note that, individual population density still remains within $[0,1]$  in region I.

\par Region II of Fig.\ \ref{fig2} (a) reveals periodic oscillatory coexistence of $\mathbf{C}$, $\mathbf{P}$ and $\mathbf{D}$ simultaneously up to the dotted vertical line. This periodic attractor demolishes through the inverse Hopf bifurcation and gives rise to stable coexisting stationary point. Here, also $y>z>x$ which establishes the dominance of punishers over other subpopulations. There are two precise differences between the region I and region II. In region I, the normalized population density $x+y+z$ is over crowded being greater than $1$. Whereas, in region II, the normalized population density lies within $(0,1)$. Secondly, region I only contains oscillatory coexistence of $x,y,z$, but segion II portrays collection of periodic attractor and stable stationary points up to $\sigma_{1}=0.775$. At the particular value of $\sigma_{1}=0.775$, we find all fraction of subpopulations are equal after the post-transient dynamics. That is, at $\sigma_{1}=0.775$, we have $x=y=z$. To distinguish this behavior with other observed phenomenon, a red dashed line is drawn in the Fig.\ \ref{fig2} (a).

\par Region III only consists stationary states. In this regime, the cooperators are dominant over other subpopulations. Till the branch point (bifurcation point) $\sigma_{1}=0.844$, the fraction of $\mathbf{C}$ is always dominating the fraction of $\mathbf{D}$, which again dominates the fraction of $\mathbf{P}$. {This behavior $x>z>y$ is also maintained for $\sigma_{1} > 0.844$, however, the fraction of punishers vanishes as $y=0$. In comparison to region II where $y>z>x$, the densities of $\mathbf{C}$ and $\mathbf{P}$ are switched in region III, where $x>z>y$ is sustained.}

\par To further validate our numerical findings, largest Lyapunov exponent of the system \eqref{7} (blue curve) is plotted by varying $\sigma_{1}$ in Fig.\ \ref{fig2} (a) using the Wolf algorithm \cite{wolf1985determining}. 
 The sign of maximal Lyapunov exponent $\lambda_{max}$ changes from $0$ to $-ve$ value, which assures the transformation of the attractor from periodic nature to stationary state of the system \eqref{7}. Clearly, the plotted maximum Lyapunov exponent agrees well with the observed bifurcation diagram in the Fig.\ \ref{fig2} (a).

\par The complex dynamics exhibited due to the interplay between
different parameters are summarized using bifurcation diagrams in Figs.\ \ref{fig2} (b)-(d). In the context of PD game dynamics, higher values of temptation parameter $\beta$ always helps to provide additional benefits to the defectors and destroys the act of cooperation. This understanding is well portrayed through the Fig.\ \ref{fig2} (b). With increasing values of $\beta$, the fraction of $\mathbf{C}$ is always decreasing up to a critical value of this parameter, and beyond that critical value, $x$ completely diminishes to $0$ leading to extinction of cooperators.
At the same critical value of $\beta$, the initial increment of punisher's population is challenged, and becomes monotonically decreasing as shown in the Fig.\ \ref{fig2} (b). This phenomenon can be well interpreted as punishers are also cooperators, and the impact of punishment is neutralized due to high temptation to the defectors. {So it is natural} that population density of punishers reduces with increasing $\beta$. {{However, punishers do not go extinct in the observed regime for $\beta \in (1,3]$, as punishment has social security in the form of reduction in the expected payoff of defectors, who need to pay an extra fine. We expect the defector population $z$ to increase with the temptation parameter $\beta$ and, therefore, find it interesting that $z$ remains constant throughout the interval $(1,3]$ of $\beta$. This may be due to the chosen values of the other parameters, which play a significant role in survivability of each subpopulation. The obtained results can also be verified using linear stability analysis (See {\mbox{\ref{Existence and stability analysis of the stationary state}}}) at the chosen values of parameters for Fig. {\mbox {\ref{fig2}}} (b).}

\par {Similarly, the role of death rate $\xi$ is inspected in Fig. {\mbox{\ref{fig2}}} (c). For the particular choice of the other parameters' values, Fig. {\mbox{\ref{fig2}}} (c) depicts that up to a certain value of $\xi$, say, $\xi_{critical} \approx 0.3605$, both the fraction of cooperators, $x$, and punishers, $y$, stay at zero, and beyond $\xi_{critical}$, both $x$ and $y$ are increasing. On the other hand, even though $z$ decreases as $\xi$ increases, $z>0$ throughout the interval $[0,0.7]$ of $\xi$.}} Hence, up to $\xi_{critical}$, the cooperator-free and punisher-free stationary point $E_{3}=\left(0,0,1-\dfrac{\xi}{\sigma_3}\right)$ is found, which is marginally stable for our particular choice of parameter values (for a detailed analysis, please see \ref{Existence and stability analysis of the stationary state}). Clearly, the $z$-component of $E_{3}$ suggests the growth of $\xi$ ultimately decreases $z$, the fraction of $\mathbf{D}$. This investigation perfectly fits with our numerical findings in Fig.\ \ref{fig2} (c). Whenever $\xi$ is greater than $\xi_{critical}$, the stationary point $E_{3}$ loses its stability, and the interior stationary point $E_{7}$ gains its stability as shown in Fig.\ \ref{fig2} (c). Thus, increment of mortality rate surprisingly encourages the coexistence of all subpopulations. {The moderate decrement in the population fraction of defectors with increasing $\xi$ substantially suppresses the defective exploitation of cooperative benefit, which may introduce a positive catalytic effect towards the concurrence of \textbf{C}, \textbf{P}, and \textbf{D}}. Even when $\xi$ is approximately close to $0.65$, the punishers dominate both $\mathbf{C}$ and $\mathbf{D}$. Note that, the initial condition for the numerical simulation is $(0.3,0.3,0.3)$, thus we do not give any additional preference, in terms of initial abundance, to the individual subpopulations. The decisive contribution of initial condition will be scrutinized in the next section.

 \begin{figure*}[ht]
	\centerline{\includegraphics[scale=0.55]{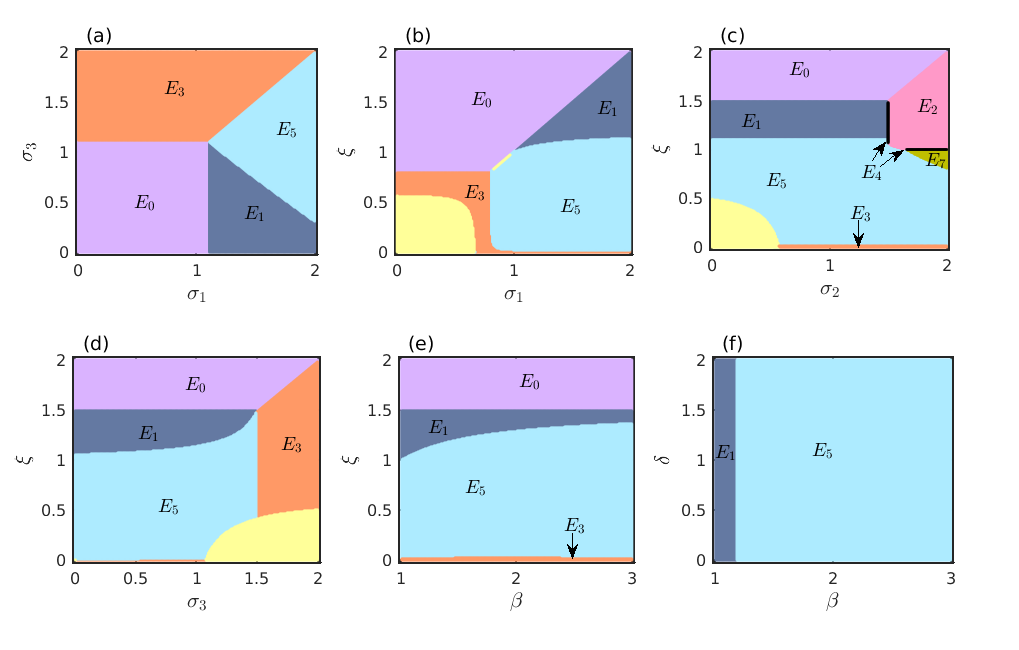}}
	\caption{{\bf Comparative plots of two-dimensional parameter phase diagrams}: {Transition between different population dependent stationary states $E_{i}$s for $i=0,1,2,\cdots,5,7$ (thoroughly addressed in Sec. {\mbox{\ref{Existence and stability analysis of the stationary state}}}) due to the interplay between the physical parameters. (a) The system dynamics under the influence of varying $\sigma_1$ and $\sigma_3$ is explored. $\sigma_1, \sigma_3 < \xi$ reveals extinction ($E_0$) of all population, $\sigma_1 < \xi$ and $\sigma_3 > \xi$ results in a defector dominant regime ($E_3$), whereas a cooperator dominant region ($E_1$) is found when $\sigma_3 < \xi$ and $\sigma_1 > \xi$. Also the choice of $\sigma_2 < \xi$ favors a punisher-free environment $E_5$. (b)-(d) Complicated reciprocity between the death rate $\xi$ and advantages given by the free spaces ($\sigma_1, \sigma_2, \sigma_3$) towards different population densities are delineated. Various combinations of population dependent steady states emerge depending on the choice of other parameter values. (e) The simultaneous contribution of $\beta$ and $\xi$ is presented. For very small (close to zero) values of $\xi$, only defectors ($E_3$) survive irrespective to the choices of $\beta$. As the death rate $\xi$ is increased gradually, the following steady states emerge in the parameter space, respectively: coexistence of cooperators and defectors $E_5$, only cooperators $E_1$, and extinction $E_0$. (f) The role of policing term $\delta$ is found to be completely independent of $\beta$ for the set of chosen parametric values. For lower values of $\beta$ only cooperators are present, but a wide range of parameter space is observed where {\mbox{{\bf D}}}s are coexisting with {\mbox{{\bf C}}}s with moderately high values of $\beta$.} Yellow region reflects unbounded solution of the system {\mbox{\eqref{7}}}. For easier comparison, whenever a pair of parameters are varied, others are fixed at these values of parameters: $\sigma_1 = 1.5$, $\sigma_2 = 0.8$, $\sigma_3 = 0.8$, $\beta = 1.2$, $\delta = 0.3$ and $\xi = 1.1$. The initial fraction of subpopulation is $(x_0, y_0, z_0)=(0.3,0.3,0.3)$ for all the figures. Lavender, cool grey, pink, coral, black, sky blue and mustard color indicate the stationary points $E_{i}$s for $i=0,1,2,\cdots,5,7$ respectively.}\label{fig3}
\end{figure*}

\par Figure \ref{fig2} (d) unveils the fundamental role of policing parameter, $\delta$. {It is clearly visible that policing of appropriate strength can fight the free-rider's defection and promote cooperation in the long run. Falling off of $z$ is evident} in  Fig.\ \ref{fig2} (d) {throughout the interval $(0,0.32]$ of $\delta$.} Up to $\delta \approx 0.2$, cooperator-free stationary point $E_{6}$ is found and at $\delta \approx 0.2$, the system \eqref{7} bifurcates and switches from $E_{6}$ to the interior equilibrium $E_{7}$ generating stable coexistence of all subpopulations. {The enhancement of punishment to the defectors marginally decreases its abundance, which in turn promotes cooperative contribution to the population via stable coexistence of all three subpopulations.} The punishers are initially enjoying the initial enhancement with the increment of $\delta$, but for $\delta > 0.2$, the growth of $\mathbf{P}$ diminishes gradually. This points out the fact that when  $\mathbf{P}$s are playing against defectors $\mathbf{D}$, to penalize them with a fine $\delta$, $\mathbf{P}$ also tolerates the cost of policing $\delta$. Thus, higher values of $\delta$ restricts the monotonically increasing nature of $\mathbf{P}$ for our chosen parameter values and initial condition. Even though the punishers are reducing in numbers with large values of $\delta$ as per this specific numerical simulation, punishment is the dominant strategy in the entire interval $(0,0.32]$ of $\delta$. It should be noted that punishment is also the dominant strategy in the Region II of Fig.\ \ref{fig2} (a), from where we choose the value of $\sigma_{1}(=0.7)$. For $\beta=1.2$ in Fig.\ \ref{fig2} (b) and $\xi=0.7$ in Fig.\ \ref{fig2} (c), punishment is the dominant strategy in our model \eqref{7}.

\par We now emphasize on the investigation of the joint impact of two parameters on the eco-evolutionary dynamics at the same time. Transitions between different stationary states are recognized due to the interplay between several physical parameters, which are used to model the system \eqref{7}. In Fig.\ \ref{fig3} (a), the system dynamics under the influence of varying $\sigma_{1}$ and $\sigma_{3}$ is explored, while the other parameters are fixed at $\sigma_2 = 0.8$, $\beta = 1.2$, $\delta = 0.3$ and $\xi = 1.1$. It should be noted that the benefits given by the free space to the punishers in the terms of positive payoff $\sigma_2=0.8$ is less than the mortality rate $\xi=1.1$ as per our chosen parameter values for this figure.  Till the free space induced reproductive benefit to the cooperators and defectors is less than their common mortality rate ($\sigma_{1}, \sigma_{3} < \xi$), Fig.\ \ref{fig3} (a) reveals extinction of all {population}. When free space gives better opportunity to any subpopulation to overcome the death rate, then that subpopulation is emerging as a dominant strategy. For instance, when $\sigma_{1} < \xi$ and $\sigma_{3} > \xi$, we notice a wide region of defector dominant regime in two-dimensional $\sigma_1-\sigma_3$ parameter space. In fact, in this regime, defectors are {the} only surviving population. 
 A reverse storyline is perceived, when $\sigma_{1} > \xi$ and $\sigma_{3} < \xi$. These extra incentives towards cooperators from $\mathbf{F}$ help to sustain cooperation and entertains a defector-punisher free, cooperator dominant region in the Fig.\ \ref{fig3} (a). As $\sigma_{2} < \xi$, our choice favors a punisher-free environment throughout the Fig.\ \ref{fig3} (a). Thus, suitable choice of all parameter values reflects a mechanism for coexistence of $\mathbf{C}$ and $\mathbf{D}$ as well.

\par The important role of death rate $\xi$ in the complex dynamics of the system \eqref{7} is now reviewed under the influence of $\sigma_1$. The equilibria $E_0$, $E_1$, $E_3$ and $E_5$ are all occurring in Fig.\ \ref{fig3} (b), similar to Fig.\ \ref{fig3} (a). A fresh captivating feature is observed over a modest region (yellow zone) in Fig.\ \ref{fig3} (b), where $\xi$ and $\sigma_1$ are both comparatively low. In this region, the variables $x$, $y$ and $z$ are leaving the phase space and {tend} to infinity after the initial transient dynamics. These types of unbounded solution are also noticed in Figs.\ \ref{fig3} (c) and (d) too. All simulations of Fig.\ \ref{fig3} are performed with fixed initial condition $(0.3,0.3,0.3)$. To understand the complicated reciprocity between the death rate and advantages given by the free spaces towards different population densities, Figs.\ \ref{fig3} (b)-(d) are delineated. If the altruist $\mathbf{F}$ is biased towards the defectors by paying them more advantages in terms of payoff $\sigma_{3} > \sigma_{1}$  and that biased favor $\sigma_{3}$ exceeds the mortality rate $\xi$, then only defector-sustaining population persists ($E_3$ in Fig.\ \ref{fig3} (b)). But, the role of other parameters like $\sigma_{2}$ and $\delta$ is also important. Depending on other parameters, cooperators are only surviving species, when $\sigma_{1} > \xi > 1$ and $\sigma_1 > \sigma_2$ in Fig.\ \ref{fig3} (b). Even, a moderate zone is noticed in Fig.\ \ref{fig3} (b), where cooperators are able to survive along with defectors. {All subpopulations go extinct}, when per capita death rate $\xi$ exceeds $\sigma_{1}, \sigma_{2}$ and $\sigma_{3}$.

\par A fascinating result is shown in Fig.\ \ref{fig3} (c), where we are able to capture distinct equilibria along with the unbounded trajectories in the 2D-parameter space of free space induced reproductive opportunity to the punishers, $\sigma_{2}$ and common mortality rate, $\xi$. For sufficiently high values of mortality rate $\xi$, all population die out. For comparatively lower values of $\xi$, the cooperators can only survive until $\sigma_{1} > \sigma_2$. For $\sigma_{1} < \sigma_2$, punishers can only survive. This transition takes place through the emergence of a small region of coexistence of $\mathbf{C}$ and $\mathbf{P}$, whenever $\sigma_1=\sigma_2$, or $\xi=1$ is satisfied. Further, lowering the values of $\xi$, defectors are found along with $\mathbf{C}$. Even, a tiny regime for large $\sigma_{2}$ is found, where coexistence of all subpopulations (the stationary state $E_7$)  occur. For too small values of $\xi$ and beyond a certain threshold of $\sigma_2$, a defector dominant solution space is obtained.

\par The important role of death rate is also demonstrated in Fig.\ \ref{fig3} (d) over the parameter free space mediated reproductive benefit to the defectors, $\sigma_{3}$. For large values of $\sigma_{3}$, either population extincts or only defectors can survive or the unbounded trajectories are found. Whereas for $\xi > \sigma_{1}, \sigma_{2}, \sigma_{3}$, extinction scenario of all species is again detected. For lower values of $\sigma_{3}$ depending on $\xi$ and other parameter values, either cooperation is the only surviving strategy, or coexistence of cooperation and defection is discovered. The simultaneous contribution of temptation parameter, $\beta$ and common death rate, $\xi$ is presented in the Fig.\ \ref{fig3} (e). Here, the parameter values are set at $\sigma_1 = 1.5$, $\sigma_2 = 0.8$, $\sigma_3 = 0.8$, and $\delta = 0.3$. For very small (close to zero) values of $\xi$, cooperator-free and punisher-free population can only be noticed irrespective to the choices of $\beta$. With increment of death rate, cooperators are coexisting with defectors, and further increment of $\xi$ demolishes the defector population. We observe an interval in the $\beta-\xi$ parameter space, where cooperators are only surviving. If $\xi$ is too high and beyond a critical threshold ($\xi > \sigma_{i}$ for $i=1,2,3$), then extinction of all species is detected.

\par Interestingly, it is expected that moderate value of the policing parameter $\delta$ always helps in persistence of punishers. But, for our chosen parameter values, Fig.\ \ref{fig3} (f) depicts a punisher-free society. With enhancing values of temptation parameter $\beta$, the defectors are getting extra aid. Thus, although initially only cooperators are present in the 2D parameter space (See Fig.\ \ref{fig3} (f)), but a wide range of parameter space is observed with moderately high values of $\beta$, where $\mathbf{D}$s are coexisting with $\mathbf{C}$. Figure \ref{fig3} (f) suggests the role of policing term $\delta$ is completely independent of $\beta$ at least for these set of chosen parametric values.

\subsection{Basin of attraction}

\begin{figure*}[ht]
	\centerline{\includegraphics[scale=0.5]{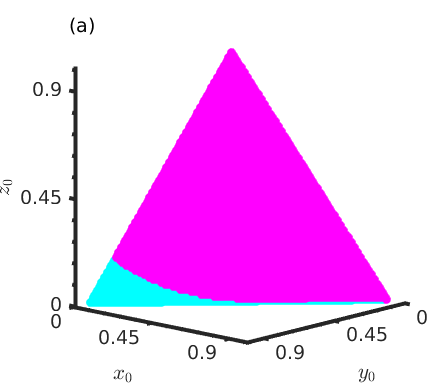}
		\includegraphics[scale=0.5]{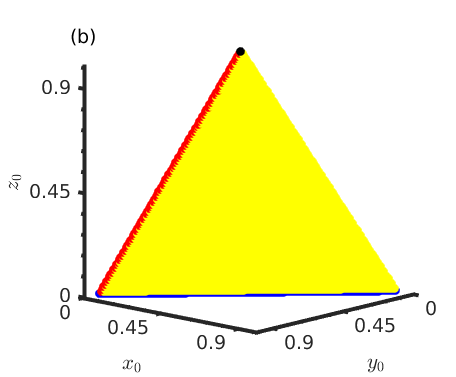}
		\includegraphics[scale=0.5]{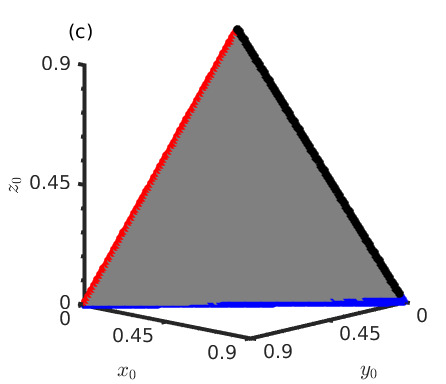}}
	\caption{{\bf Basin of attraction of system \eqref{7} in 3D $xyz$-plane}: Three distinct sets of parameter values are chosen to demonstrate the effect of initial conditions $(x_0,y_0,z_0)$. $x_0$, $y_0$ and $z_0$ are varying within $[0,0.9]$, while the relation $x_0+y_0+z_0=0.9$ is always kept intact. {The system converges to various attractors, or grows without any bound solely depending upon the initial abundance of each population with fixed parameter values. (a) Two distinct cooperator-free stationary states $E_2$ and $E_3$ are obtained. (b)-(c) Three different kinds of stationary points are found where the boundaries of the basin of attraction lead to different stationary points, compared to the interior of the basin.} Colors: Cyan $\to$ $E_2$, Magenta $\to$ $E_3$, Blue $\to$ Unbounded trajectory, Red $\to$ $E_6$, Yellow $\to$ $E_5$, Black $\to$ $E_0$, Gray $\to$ $E_7$. The meaning of these stationary points $E_i$, $i=0,1,2,\cdots,6$ are addressed in {\mbox{\ref{Existence and stability analysis of the stationary state}}}. Parameter values: (a) $\sigma_1 = 1.2$, $\sigma_2 = 1.5$, $\sigma_3 = 1.4$, $\beta = 1.5$, $\delta = 0.5$ and $\xi = 1.1$; (b) $\sigma_1 = 1.2$, $\sigma_2 = 1.0$, $\sigma_3 = 0.41$, $\beta = 2.6$, $\delta = 1.39$ and $\xi = 0.5$; and (c) $\sigma_1 = 0.51$, $\sigma_2 = 1.0$, $\sigma_3 = 0.41$, $\beta = 1.1$, $\delta = 0.1$ and $\xi = 0.7$. }\label{fig4}
\end{figure*}

\par Initially with the help of Fig.\ \ref{fig1}, we have discussed that multistability is observed in our model. Depending on the initial condition, the state converges either to different Nash equilibria consisting of cooperators, defectors and punishers, or to periodic orbits, where the frequencies of punishers, defectors, and cooperators oscillate endlessly. To illustrate this feature, basin of attraction for three discrete sets of parameters is shown in Fig.\ \ref{fig4}. We set $w_0=0.1$ for all of these subfigures, so that each of the variables $x_0$, $y_0$, $z_0$ varies within the interval $[0,0.9]$ maintaining the relation $x_0+y_0+z_0=0.9$. Each subfigure in Fig.\ \ref{fig4} depicts that the system \eqref{7} converges to various attractors, or grows without any bound solely depending upon the initial abundance of each subpopulation with fixed parameter values. In Fig.\ \ref{fig4} (a), two distinct cooperator-free stationary states are obtained. In this particular figure, we set the values of the parameters at $\sigma_1 = 1.2$, $\sigma_2 = 1.5$, $\sigma_3 = 1.4$, $\beta = 1.5$, $\delta = 0.5$ and $\xi = 1.1$. These values satisfy local stability criterion for both stationary points $E_2$ (cyan) and $E_3$ (magenta) respectively (See \ref{Existence and stability analysis of the stationary state}). The chosen parameter set establishes that the free space induced benefit to the cooperators is the least as compared to the free space mediated benefits to the punishers and defectors ($\sigma_1<\sigma_2,\sigma_3$). 
	 The Fig.\ \ref{fig4} (a) reveals that with suitable choice of initial fraction of each subpopulation, one can reach any one of these cooperator-free stable nodes $E_2$ and $E_3$.

\par Similarly, three different stationary points are obtained in Fig.\ \ref{fig4} (b). Surprisingly, the boundaries of the basin of attraction lead to different stationary points, compared to the interior of the basin. The interior region of the basin of attraction helps the system \eqref{7} to reach the punisher-free stable steady state $E_5$. {We even find a single initial condition} $(x_0,y_0,z_0)=(0,0,0.9)$ located at the top vertex of the triangle, for which the system \eqref{7} converges to the extinction stationary point $E_0$ (black). The general solution (flow) with real-valued expansion coefficients for the system \eqref{7} is given by

\begin{equation} \label{10}
\begin{array}{lcl}
d_1 e^{\lambda_{1} t} \mathbf{u_1}+d_2 e^{\lambda_{2} t} \mathbf{u_2}+d_3 e^{\lambda_{3} t} \mathbf{u_3},
\end{array}
\end{equation}
where $\mathbf{u_i}$'s are the eigen vectors corresponding to the eigen values $\lambda_{i}$ of Jacobian $J$ of the linearized system for the chosen fixed set of parameter values for $i=1,2,3$. The $\lambda_{i}$s are explicitly calculated at the $J(E_0)$ in \ref{Existence and stability analysis of the stationary state}. The initial conditions with $x_0=0$ and $y_0=0$ yield the constants $d_1=0=d_2$ and hence, the solutions tend to the stationary point at the origin as $\lambda_{3} < 0$. An elaborate discussion regarding the role of these boundary initial conditions with at least one zero-component is rigorously addressed in \ref{Analysis of the system}. There is a line of initial conditions (blue) on the lower boundary of basin of attraction, which yields diverging orbits. Along these initial conditions, the constant $d_3$ is zero. As for our chosen parameter values $\sigma_1 = 1.2$, $\sigma_2 = 1.0$, $\sigma_3 = 0.41$, $\beta = 2.6$ and $\delta = 1.39$, $\lambda_{1}, \lambda_{2} > 0$ of the Jacobian $J$ at $E_0$ and thus, Eqn.\ \eqref{10} tends to infinity {in the long run} (as $t \to \infty$). 
 Another line of initial conditions on the left boundary (red) converges to cooperator extinction equilibrium $E_6$.

\par Stable coexistence of all subpopulations is observed in Fig.\ \ref{fig4} (c) for the chosen fixed values of parameters $\sigma_1 = 0.51$, $\sigma_2 = 1.0$, $\sigma_3 = 0.41$, $\beta = 1.1$, $\delta = 0.1$ and $\xi = 0.7$. The boundary of the set of initial conditions converges to either of two different stable stationary points; the extinction of population, $E_0$ (black) and the stable concurrence between punishers and defectors, $E_6$ (red). The interior of the basin helps to obtain the stable coexistence of all three subpopulations, $E_7$ (gray). There exists a few initial conditions for which unbounded trajectories (blue) are the only possible solutions. Initial conditions with $y_0=0$ lead to $d_2=0$ in Eq.\ \eqref{10} and $\lambda_{1}, \lambda_{3}$ of $J(E_0)$ will be negative for our chosen parameter values. Thus the general solution converges to the stationary point $E_0$ for the initial conditions on the right boundary $y_0=0$ of the basin of attraction.

\section{Conclusion: Summary and final comments }

\par The influence of ecology on the evolution of population (eco-to-evo) and inversely, the impact of population's evolution on ecology (evo-to-eco) encourage a lot of young researchers to focus on how change in one process affects the change on the other. On the other hand, the evolution of rational behavior among population is ideally described using evolutionary game dynamics. This primarily inspects how cooperation emerges inside a population community by overcoming the social dilemma of what is the best for own and what is the best for the society. We look over this Darwinian puzzle by integrating ecologically-accessible free space with the evolution of population in the framework of evolutionary game theory. For this purpose, we consider PD game, in particular, as a paradigm for tackling the problem of cooperation. The game promises that defection always results in a better payoff than cooperation, and thus, two independent rational individuals might defect each other, even if cooperation is the best choice for the group.

\par A defecting individual always receives the highest fitness if facing a cooperator. To solve this riddle from the evolutionary viewpoint, a new strategy punishment is adopted. Punishers pay a cost to punish the defectors. {Punishment has been found as one of the emergent spontaneous behaviors of the human society as a way of treating the defectors for their free-riding mentality.} Many previous theoretical works \cite{brandt2006punishing,wang2013impact,banerjee2019delayed,helbing2010defector,perc2012sustainable,dreber2008winners,egas2008economics,sasaki2007probabilistic, henrich2001people,bowles2004evolution,ohtsuki2009indirect,hauert2002replicator,brandt2006good,liu2018evolutionary} have revealed the role of punishment for the better understanding of cooperation. However, studies related to the combined effect of altruistic act of free space towards providing the reproductive benefit to the constituents and the punishment are relatively missing in the existing literature to the best of our knowledge. In this paper, we have introduced four distinct competing strategies, viz.\ cooperation, punishment, defection and free spaces. The interplay of these strategies are particularly common and relevant in our real society. The strategy free space does not take any advantages for providing the benefits to other subpopulations. In fact, any individual from the subpopulation $\mathbf{C}$, $\mathbf{P}$ and $\mathbf{D}$ can use free space for their replication. In order to shed some light on this one-sided contribution of free space, we have constructed a general mathematical model by combining game and  ecological dynamics, where the interaction pattern between cooperators and defectors follows the contemporary PD game. {This type of selfless, altruistic act \mbox{\cite{axelrod1984evolution}} can commonly be observed in ants, bacteria, birds, bees, and many other higher mammals. Our eco-evolutionary model captures this remarkable aspect of biological and behavioral sciences \mbox{\cite{nowak2006five}} using the selfless act of free space, which makes an effort to improve the welfare of others by sacrificing personal benefits.} 


\par 
 The model developed in this study consists three variables and six parameters. These parameters have important implications
in many settings of ecological network, infectious disease dynamics, animal behavior and social interactions of humans. The different choices of these parameter values lead to several emergent dynamics, and our numerical investigations are restricted to only finite possible alternatives of this uncountable parameter space. However, we are able to capture the essence of the multistable replicator dynamics for various possible values of different physical parameters. The model studied here corresponds to scenarios in which cyclic dominance can be maintained through the occurrence of periodic attractor. {Such kind of cyclical interaction \mbox{\cite{szolnoki2014cyclic}} is capable of capturing the beauty of the governing eco-evolutionary dynamics, and similar behavior is found to occur in many real-life instances including the mating strategy of side-blotched lizards, the genetic regulation in the repressilator, the overgrowth of marine sessile organisms, competition in microbial populations, and many more.} Few snapshots of such periodic attractor and their temporal evolution are shown for several parameter values. Even, for a particular set of parameter values, we are able to demonstrate the inverse Hopf route for destruction of these periodic attractors. The result is also validated using the largest Lyapunov exponent of the system \eqref{7}. {This Hopf bifurcation, yielding periodic oscillations through destabilization of the steady state behavior, is ubiquitous in many biological and physical systems including Lotka-Volterra model of predator-prey interaction, the Lorenz attractor, the Selkov model of glycolysis, the Hodgkin-Huxley model for nerve membranes, to name but a few examples.} Interestingly, slow-fast time scales are noticed for our model \eqref{7} during the manifestation of such periodic attractor. This periodic orbit gives all species a fair chance to dominate one another in a cyclic fashion.

\par We have also been able to map the different potential dynamical states in the two-dimensional parameter plane by keeping fixed the other four parameter values. Various stationary states  are obtained during numerical investigation reflecting five different possibilities: (i) extinction of all subpopulations, (ii) existence of only one outcompeting subpopulation, (iii) survival of any two subpopulations, (iv) coexistence of all subpopulations, and (v) unbounded diverging orbits. The reasoning behind these results are thoroughly addressed using physical interpretations of all parameters. The understanding is further explicated using linear stability analysis of the eco-evolutionary dynamics.



\par From our analytical findings and associated numerical simulation results, it is clear that if the mortality rate is higher than the benefits provided by the free space, then it is almost impossible for the species to survive. Our results unveil the influence of the death rate, which proves to be quite significant in maintaining biodiversity. {With suitable contribution of other parameters, increment of mortality rate as well as the policing parameter is found to encourage the coexistence of all subpopulations (Figs \mbox{\ref{fig2}} (c) and \mbox{\ref{fig2}} (d)). The contribution of the temptation parameter in PD game, that disrupts the evolution of cooperative nature of individuals by providing greater benefit to defectors, is well established in the literature. Consistently, if the temptation to defect is sufficiently large, our approach may fail to sustain cooperation, as illustrated in Fig. \mbox{\ref{fig2}} (b). Even, this figure contemplates the decreasing fraction of punishers with increment of temptation parameter $\beta$. In addition, potential evolutionary advantage of punishment is presented in Fig. \mbox{\ref{fig2}} (d). Suitable choice of policing parameter $\delta$ helps to maintain the survivability of both the traditional cooperators as well as the punishers (that are also cooperative in nature), thereby restricting the total extinction of cooperators.  Figure \mbox{\ref{fig2}} (c) reveals the fascinating twist that the increment of mortality rate $\xi$ ultimately leads to the collapse of defector's population, and consequently, coexistence of all subpopulations under favourable conditions is observed. In fact, if free space is biased towards a particular subpopulation and that free space induced advantage is higher than the common mortality rate, then our eco-evolutionary model may help to promote that particular subpopulation. This behavior is demonstrated in Fig. \mbox{\ref{fig3}}. For instance, if cooperation is favoured by the free space compared to other strategies (i.e., $\sigma_1 > \sigma_{2},\sigma_{3}$) and this favouritism $\sigma_{1}$ is higher than the death rate $\xi$, then only cooperators survive and other subpopulations become extinct (See the cool grey region of Fig. \mbox{\ref{fig3}} (a)). Moreover, the observed phenomenon of multistability that reveals coexistence of more than one attractors is also emphasized in detail throughout the article. Figures \mbox{\ref{fig1}} and \mbox{\ref{fig4}} ensuring the multistable dynamics exhibited by our model points out the vulnerability of the system to small perturbations. The presence of multistability and multiple operating regimes are essential for biology such as in prey-predator commuinities, biochemical responses and generation of cell cycle oscillation \mbox{\cite{angeli2004detection,banerjee2020cooperative}}.} To clarify the understanding behind the multistability, particularly at the boundaries of the basin of attraction, mathematical analysis is found to be effective.

\par In conclusion, our constructed model provides certain features with several significant feasible inferences. Our study supports a deeper understanding of the impact of free space induced reproductive benefit on the evolution of population, where the act of punishment improves the emergence and promotion of population-wide cooperation. It comes up with an effective yet simple way for the promotion of the stable coexistence of different strategies including cooperation, which  may lead to an interesting direction for future research and for better understanding of the ecological balance in nature.

\section*{CRediT authorship contribution statement}

\par  {\bf Sayantan Nag Chowdhury \& Srilena Kundu}: Conceptualization; Methodology; Software; Validation; Formal analysis; Investigation; Writing - original draft; Visualization; {\bf Jeet Banerjee}: Conceptualization; Validation; Methodology; Visualization; Writing - review \& editing; {\bf Matja{\v z} Perc}: Validation; Visualization; Writing - review \& editing; {\bf Dibakar Ghosh}: Supervision; Validation; Visualization; Writing - review \& editing.

\section*{Acknowledgement}
{The authors gratefully acknowledge the anonymous referees for their insightful suggestions that helped in considerably improving the manuscript. }SNC and DG were supported by the Department of Science and Technology, Government of India (Project No. EMR/2016/001039). SNC would also like to acknowledge the financial support from Indian Statistical Institute, Kolkata and the CSIR  (Project No. 09/093(0194)/2020-EMR-I) for funding him during the end part of this work. MP was supported by the Slovenian Research Agency (Grant Nos. P1-0403, J1-2457, J4-9302, and J1-9112).

\appendix
\section{Existence, uniqueness and positive invariance of solutions} \label{Positivity and boundedness of solutions}



\par Positivity of a model guarantees that the model is biologically well behaved. It is easy to notice that the functions on the right side of each of the equations of system \eqref{7} are continuously differentiable in $\mathbb{R} \times \mathbb{R} \times \mathbb{R}$. Thus, the solution of Eqs. \eqref{7} with a positive initial condition always exists. Also, the uniqueness of solutions for the system \eqref{7} in $\mathbb{R}^{3}_{+}$ is assured, as the right-hand side of each of the equations in system \eqref{7} is locally Lipschitz in the first quadrant. The solution of system \eqref{7} in terms of time $t \ge 0$ can be written in the form

\begin{equation} \label{8}
\begin{array}{lcl}
x(t) =x(0) \exp\left[\int_{0}^{t} \phi_1(x,y,z,\sigma_1,\xi) ds \right],\\\\
y(t) =y(0) \exp\left[\int_{0}^{t} \phi_2(x,y,z,\sigma_2,\delta,\xi) ds \right],\\\\
z(t) =z(0)\exp\left[\int_{0}^{t} \phi_3(x,y,z,\sigma_3,\delta,\xi,\beta) ds \right],
\end{array}
\end{equation}

where,
\begin{equation} \label{9}
\begin{array}{lcl}
\phi_1(x,y,z,\sigma_1,\xi)=(1-\sigma_1)x+(1-\sigma_1)y-\sigma_1 z  \\~~~~~~~~~~~~~~~~~~~~~~~~~+(\sigma_1-\xi),\\ \phi_2(x,y,z,\sigma_2,\delta,\xi)=(1-\sigma_2)x+(1-\sigma_2)y-(\sigma_2+\delta)z \\~~~~~~~~~~~~~~~~~~~~~~~~~~~~~+(\sigma_2-\xi), \\ \phi_3(x,y,z,\sigma_3,\delta,\xi,\beta)=(\beta-\sigma_3)x+(\beta-\sigma_3-\delta)y-\sigma_3 z \\~~~~~~~~~~~~~~~~~~~~~~~~~~~~~~~~~+(\sigma_3-\xi).
\end{array}
\end{equation}

\par The system of integral equations \eqref{8} asserts all the solutions of the system \eqref{7} that start in $\mathbb{R}^{3}_{+}$ remain positive for all the time.



\section{Existence and stability analysis of the stationary state} \label{Existence and stability analysis of the stationary state}

\subsection{Stationary states and their existence}

\par Setting $\dfrac{dx}{dt}=0$, $\dfrac{dy}{dt}=0$, and $\dfrac{dz}{dt}=0$, the system \eqref{7} has at most eight non-negative equilibria, viz.\

\begin{enumerate}
	\item The trivial extinction stationary point $E_{0}=(0,0,0)$.
	\item The punisher-free and defector-free stationary point $E_{1}=\left(\dfrac{\sigma_1-\xi}{\sigma_1-1},0,0\right)$. This stationary point exists, i.e., only cooperators are present if $\sigma_1 > \xi \ge 1$, or $0 < \sigma_1 < \xi \le 1$.
	\item The cooperator-free and defector-free stationary point $E_{2}=\left(0,\dfrac{\sigma_2-\xi}{\sigma_2-1},0\right)$, which exists if $\sigma_2 > \xi \ge 1$, or $0 < \sigma_2 < \xi \le 1$.
	\item The cooperator-free and punisher-free stationary point $E_{3}=\left(0,0,1-\dfrac{\xi}{\sigma_3}\right)$. In this case, only defector exists, if $\sigma_{3} > \xi$.
	\item The defector-free stationary point $E_{4}=\left(\alpha_1,\alpha_2,0\right)$, where $\alpha_1+\alpha_2=\dfrac{\sigma_1-\xi}{\sigma_1-1}$. This stationary point exists, if $\sigma_1 > \xi > 1$ or $0 < \sigma_1 < \xi < 1$ and $\sigma_1=\sigma_2$. If $\xi=1$, then $\sigma_{1} (\ne 1)$ need not be equal to $\sigma_{2} (\ne 1)$ for existence of $E_{4}$, where $\alpha_1+\alpha_2=1$.
	
	\item The punisher-free stationary point $E_{5}=\left(\eta_{1},0, 1-\eta_{1}+\eta_{2} \right)$, where
	$\eta_{1}=\dfrac{\xi(\sigma_1-\sigma_3)}{\beta \sigma_1-\sigma_3}$ and $\eta_{2}=\dfrac{\xi(1-\beta)}{\beta \sigma_1-\sigma_3}$. Clearly, $\eta_{1}$ lies within $(0, 1)$, if $(\xi-\beta)\sigma_1 < (\xi-1)\sigma_3$ for $\beta \sigma_1 - \sigma_3 > 0$ or, $(\xi-\beta)\sigma_1 > (\xi-1)\sigma_3$ when $\beta \sigma_1 - \sigma_3 < 0$. Similarly, $1-\eta_{1}+\eta_{2} < 1$, if $(1-\beta) < (\sigma_1-\sigma_3)$ for $\beta \sigma_1 - \sigma_3 > 0$ and $\xi >0$ or, $(1-\beta) > (\sigma_1-\sigma_3)$ for $\beta \sigma_1 - \sigma_3 < 0$ and $\xi > 0$. The $z$-component will be positive, if $\beta \sigma_1 - \sigma_3 < \xi (\sigma_1 - \sigma_3 -1 + \beta)$ for $\beta \sigma_1 - \sigma_3 < 0$, or, $\beta \sigma_1 - \sigma_3 > \xi (\sigma_1 - \sigma_3 -1 + \beta)$ for $\beta \sigma_1 - \sigma_3 > 0$. Also, $-1 < \eta_{2} \le 0$, as the sum of $x$ and $z$ components should be bounded above by $1$ and bounded below by $0$.
	
	\item The cooperator-free stationary point $E_{6}=\left(0, \gamma_1, 1- \gamma_1 + \gamma_2 \right)$, where  $\gamma_1=\dfrac{\xi (\sigma_{2}-\sigma_{3})+\delta (\xi-\sigma_{3})}{\Delta}$, $\gamma_2=\dfrac{\xi (1-\beta +2\delta)-\delta(\beta-\delta)}{\Delta}$ and $\Delta=(\beta \sigma_2 -\sigma_3)+\delta(\beta-\delta-\sigma_2-\sigma_3) \ne 0$. This stationary point exists, if
	\[ \begin{cases}
	\delta \sigma_{3} < \xi(\sigma_{2}-\sigma_{3}+\delta) < (\beta-\delta)(\delta+\sigma_{2})-\sigma_{3} \\
	\text{and}\\
	\delta(\delta+\sigma_{3}-\beta) < \xi (\sigma_{2}-\sigma_{3}-\delta-1+\beta)\\
	<\beta \sigma_{2}-\sigma_{2} \delta -\sigma_{3} \\
	 \text{for} \hspace{0.5cm} \Delta > 0
	\end{cases}
	\]
	or,
	\[ \begin{cases}
	\delta \sigma_{3} > \xi(\sigma_{2}-\sigma_{3}+\delta) > (\beta-\delta)(\delta+\sigma_{2})-\sigma_{3} \\
	\text{and}\\
	\delta(\delta+\sigma_{3}-\beta) > \xi (\sigma_{2}-\sigma_{3}-\delta-1+\beta)\\
	> \beta \sigma_{2}-\sigma_{2} \delta -\sigma_{3} \\
	\text{for} \hspace{0.5cm} \Delta < 0.	
	\end{cases}
	\]
	Also, $-1 < \gamma_{2} \le 0$ should be hold.
	
	\item 

	The interior stationary point $E_{7}=(\zeta_{1}, \zeta_{2}, \zeta_{3})$, where
	\[
	\begin{cases}
	\zeta_{1}=-\gamma-\alpha-\zeta_{3}+\Delta_1,\\
	\zeta_{2}=\alpha+\gamma,\\
	\zeta_{3}=\dfrac{(\xi-1)(\sigma_{2}-\sigma_{1})}{\sigma_{1}-\sigma_{2}-\delta+\sigma_{1}\delta},\\
	\alpha=\dfrac{(1-\xi)(\sigma_1-\sigma_3)}{\sigma_{1}-\sigma_{2}-\delta+\sigma_{1}\delta},\\
	\gamma=\dfrac{(\beta-1)[\xi(\sigma_1-\sigma_2-\delta) +\sigma_1\delta]}{\sigma_{1}-\sigma_{2}-\delta+\sigma_{1}\delta},\\
	\Delta_1=1+\dfrac{\delta(1-\xi)}{\sigma_{1}-\sigma_{2}-\delta+\sigma_{1}\delta},\\
	\text{and} \hspace{0.2cm} \sigma_{1}-\sigma_{2}-\delta+\sigma_{1}\delta \ne 0.
	\end{cases}
	\]

	 This stationary point exists, if $ 0 < \zeta_{1}, \zeta_{2}, \zeta_{3} <1$, $0<\zeta_{1}+\zeta_{2}+\zeta_{3} \le 1$ and
	\[\begin{cases}
	\xi(\sigma_{2}-\sigma_{1}) < \delta(\sigma_{1}-1)\\
	(\xi-1)(\sigma_{2}-\sigma_{1})>0\\
	\text{for} \hspace{0.5cm} \sigma_{1}-\sigma_{2}-\delta+\sigma_{1}\delta > 0,
	\end{cases}
	\]
	or,
	\[\begin{cases}
	\xi(\sigma_{2}-\sigma_{1}) > \delta(\sigma_{1}-1)\\
	(\xi-1)(\sigma_{2}-\sigma_{1})<0\\
	\text{for} \hspace{0.5cm} \sigma_{1}-\sigma_{2}-\delta+\sigma_{1}\delta < 0.
	\end{cases}
	\]
		
\end{enumerate}	

\subsection{Stationary states and their local stability}

\par The Jacobian matrix of the system \eqref{7} at any stationary point $(x_{*}, y_{*}, z_{*})$ can be expressed as

 $ J(x_{*}, y_{*}, z_{*})=\left( \begin{array}{ccccc}
J_{11} & J_{12} & J_{13}\\
J_{21} & J_{22} & J_{23}\\
J_{31}  & J_{32} & J_{33}
\end{array} \right)$,

where
\[\begin{cases}
J_{11}=2(1-\sigma_{1})x_{*}+(1-\sigma_{1})y_{*}-\sigma_{1}z_{*}+(\sigma_{1}-\xi),\\
J_{12}=(1-\sigma_{1})x_{*}\\
J_{13}=-\sigma_{1}x_{*},\\
J_{21}=(1-\sigma_{2})y_{*},\\
J_{22}=(1-\sigma_{2})x_{*}+2(1-\sigma_{2})y_{*}-(\sigma_{2}+\delta)z_{*}+(\sigma_{2}-\xi),\\
J_{23}=-(\sigma_{2}+\delta)y_{*},\\
J_{31}=(\beta-\sigma_{3})z_{*},\\
J_{32}=(\beta-\sigma_{3}-\delta)z_{*},\\
J_{33}=(\beta-\sigma_{3})x_{*}+(\beta-\sigma_{3}-\delta)y_{*}-2\sigma_{3} z_{*}+(\sigma_{3}-\xi).
\end{cases}\]

\par The different equilibria of the system and their stability properties are described below:

\begin{enumerate}
	\item The trivial equilibrium $E_{0}$ is asymptotically stable node, if $\sigma_1$, $\sigma_2$, $\sigma_{3} < \xi$ with $\xi > 0$. The eigenvalues $\lambda_{i}$ of the Jacobian matrix $J$, evaluated at $E_{0}$ are given by $\lambda_i = \sigma_i -\xi$, $i = 1, 2, 3$.
	
	
	\item The eigenvalues of $J(E_{1})$ are
	  \[
	  \begin{cases}
	  \lambda_{1}=\xi-\sigma_{1},\\
	  \lambda_{2}=\dfrac{(\sigma_{1}-\sigma_{2})(1-\xi)}{\sigma_1-1},\\
	  \lambda_{3}=\dfrac{\beta \sigma_{1}-\sigma_{3}+\xi(1-\beta+\sigma_3-\sigma_1)}{\sigma_1-1}.
	  \end{cases}
	  \]
	  \par Thus, the stability criteria of the node $E_{1}$ reduces to $\sigma_{1} > \sigma_{2},\ \sigma_{1} > \xi >1$ and $\beta \sigma_{1}-\sigma_{3}+\xi(1-\beta+\sigma_3-\sigma_1) < 0$.
	
	
	  \item The eigenvalues of the Jacobian $J$ at the stationary point $E_{2}$ are
	  \[
	  \begin{cases}
	  \lambda_{1}=\dfrac{(1-\xi)(\sigma_{2}-\sigma_{1})}{\sigma_{2}-1},\\
	  \lambda_{2}=\xi-\sigma_{2},\\
	  \lambda_{3}=\dfrac{\xi(1-\sigma_{2}+\sigma_{3}-\beta+\delta)+\sigma_{2}(\beta-\delta)-\sigma_{3}}{\sigma_{2}-1}.
	  \end{cases}
	  \]
	
	  The negative values of these set of eigenvalues suggest $E_{2}$ is a stable node. Thus, the solution of system \eqref{7} containing only punishers ($y$) is stable, if $\sigma_{2} > \sigma_{1},\ \sigma_{2} > \xi > 1$ and $\xi(1-\sigma_{2}+\sigma_{3}-\beta+\delta)+\sigma_{2}(\beta-\delta)-\sigma_{3} < 0$.
	
	
	  \item The eigenvalues of $J(E_{3})$ are
	  \[
	  \begin{cases}
	  \lambda_{1}=\xi(\dfrac{\sigma_{1}}{\sigma_{3}}-1),\\
	  \lambda_{2}=\dfrac{\xi(\sigma_{2}+\delta-\sigma_{3})-\delta\sigma_{3}}{\sigma_{3}},\\
	  \lambda_{3}=\xi-\sigma_{3}.
	  \end{cases}
	  \]
	  Hence, $E_{3}$ will be a stable node, if $0<\xi<\sigma_{3},\ \sigma_{3} > \sigma_{1}$ and $\delta \sigma_{3} > \xi (\sigma_{2}+\delta-\sigma_{3})$.
	
	
	  \item The eigenvalues of the jacobian $J$ at the stationary point $E_{4}=(\alpha_1,\alpha_2,0)$ are
	  \[
	  \begin{cases}
	  \lambda_{1}=(\beta-\sigma_{3})\dfrac{\xi-\sigma_{1}}{1-\sigma_{1}}-\delta \alpha_2 + (\sigma_{3}-\xi),\\
	  \lambda_{2}=0,\\
	  \lambda_{3}=\xi-\sigma_{1}.
	  \end{cases}
	  \]
	
	  Note that, $\alpha_1$ and $\alpha_2$ are related by the relation $\alpha_1+\alpha_2=\dfrac{\xi-\sigma_{1}}{1-\sigma_{1}}$ with $\sigma_{1} \ne 1$. Therefore, the stationary point $E_{4}$ is marginally stable, if $\xi < \sigma_{1}$ and $\lambda_{1} < 0$.

\par Till now, using the existential criterion and the negativity of the eigenvalues, the stability of the autonomous system \eqref{7} is investigated. But, the eigenvalues of the Jacobian matrix $J$ at the remaining stationary points $E_5$, $E_6$ and $E_7$ are very complicated to work out. Depending on the stationary points $E_5$, $E_6$ and $E_7$ and the various parameters, the roots of the complex polynomials possess at least one real eigenvalue and the remaining two characteristic roots may be complex conjugate or real, solely depending on the different values of parameters and stationary points $E_5$, $E_6$ and $E_7$.
	

       \item Using Routh-Hurwitz stability criterion, $E_{5}=(\eta_{1},0,1-\eta_{1}+\eta_{2})$ is stable, if
       \[
       \begin{cases}
       \eta_{1}-\sigma_{2} \eta_{2}-\xi-\delta(1+\eta_{2}-\eta_{1})<0, \\
       \eta_{1} (\beta + \sigma_{3} +2 - \sigma_{1})-\sigma_{3} (1+2\eta_{2})-\sigma_1 \eta_{2} - 2\xi < 0, \hspace{0.2cm}  \text{and}\\
       \big[ 2\eta_{1} - \sigma_{1}(\eta_{1}+\eta_{2})-\xi \big] \big[ \beta \eta_{1} - \xi + \sigma_{3} (\eta_{1} -1 -2\eta_{2}) \big]\\
       +\sigma_{1} \eta_{1} (\beta-\sigma_{3})(1+\eta_{2}-\eta_{1})>0,
       \end{cases}
       \]

       \item The cooperator-free stationary point $E_{6}=\left(0, \gamma_1, 1- \gamma_1 + \gamma_2 \right)$ is stable, if
       \[
       \begin{cases}
       \gamma_1-\sigma_{1} \gamma_2 -\xi < 0,\\
       \gamma_1 (\beta +\sigma_{3}+2-\sigma_{2})-\sigma_{3}(1+2\gamma_2)\\-\gamma_2(\sigma_2+\delta)-2\xi-\delta<0,\\
       \text{and} \hspace{0.2cm}  \big[2(1-\sigma_{2})\gamma_{1}+(\sigma_2-\xi)\big]\big[(\beta-\sigma_{3}-\delta)\gamma_1\\-2\sigma_{3}(1-\gamma_1+\gamma_2)+(\sigma_{3}-\xi)\big]\\+(\sigma_{2}+\delta)(1-\gamma_1+\gamma_2)\big[2\sigma_{3}(1-\gamma_1+\gamma_2)\\-(\sigma_{3}-\xi)\big]>0,
       \end{cases}
       \]

       \item 
       Routh-Hurwitz stability criterion yields that $E_7$ is stable, if
       \[
       \begin{cases}
       a_{11}+a_{22}+a_{33}<0,\\
       a_{11}a_{23}a_{32}-a_{11}a_{22}a_{33}+a_{12}a_{21}a_{33}\\-a_{12}a_{23}a_{31}-a_{13}a_{32}a_{21}+a_{13}a_{31}a_{22}>0,\\
       -(a_{11}+a_{22}+a_{33})(a_{11}a_{22}+a_{11}a_{33}+a_{22}a_{33}\\-a_{23}a_{32}-a_{12}a_{21}-a_{13}a_{31})> (-a_{11}a_{22}a_{33}\\+a_{11}a_{23}a_{32}+a_{12}a_{21}a_{33}-a_{12}a_{23}a_{31}\\-a_{13}a_{32}a_{21}+a_{13}a_{31}a_{22}),
       \end{cases}
       \]
where,

        \[
       \begin{cases}
       a_{11}=2(1-\sigma_{1})(\Delta_{1}-\alpha-\gamma-\zeta_{3})\\+(1-\sigma_{1})(\alpha+\gamma)-\sigma_{1}\zeta_{3}+(\sigma_{1}-\xi),\\
       a_{12}=(1-\sigma_{1})(\Delta_{1}-\alpha-\gamma-\zeta_{3}),\\
       a_{13}=-\sigma_{1} (\Delta_{1}-\alpha-\gamma-\zeta_{3}),\\
       a_{21}=(1-\sigma_{2})(\alpha+\gamma),\\
       a_{22}=(1-\sigma_{2})(\Delta_{1}-\alpha-\gamma-\zeta_{3})\\+2(1-\sigma_{2})(\alpha+\gamma)-(\sigma_{2}+\delta)\zeta_{3}\\+(\sigma_{2}-\xi),\\
       a_{23}=-(\sigma_{2}+\delta)(\alpha+\gamma),\\
       a_{31}=(\beta-\sigma_{3})\zeta_{3},\\
       a_{32}=(\beta-\sigma_{3}-\delta)\zeta_{3}, \hspace{0.5cm} \text{and}\\
       a_{33}=(\beta-\sigma_{3})(\Delta_{1}-\alpha-\gamma-\zeta_{3})\\+(\beta-\sigma_{3}-\delta)(\alpha+\gamma)-2\sigma_{3}\zeta_{3}+(\sigma_{3}-\xi).
       \end{cases}
       \]
	
\end{enumerate}

\subsection{Analysis of the system \eqref{7}, when at least one initial component of $(x_{0}, y_{0}, z_{0})$ is zero} \label{Analysis of the system}

\begin{enumerate}
	\item If $(x_{0}, y_{0}, z_{0})=(0,0,0)$, then the system \eqref{7} will always converge to $E_{0}$ irrespective choice of any parameters, since $(0,0,0)$ is a fixed point of the functions on the right-hand side of each of the Eqs.\ \eqref{7}. Physically this result will be meaningful in the sense that there will be no entertainment of replication, if there are no species available in the society initially.
	
	\item If $x_{0}=0$ and $y_{0}=0$, then the system \eqref{7} is exactly solvable and the component of $x,y$ will be $0$ for all remaining time $t$, and
	
	$z=\dfrac{(\sigma_{3}-\xi)\big[1+\tanh(c_1+t)\big(\frac{\sigma_{3}}{2}-\frac{\xi}{2}\big)\big]}{2\sigma_{3}} \hspace{0.2cm} \text{with} \hspace{0.2cm} \sigma_{3} \ne 0$, where $c_1$ is a constant depending on initial condition $z_0$. This kind of initial condition may prefer defector dominated alliance within the system depending on the values of $\xi$ and $\sigma_3$.
	
	\item If $x_{0}=0$ and $z_{0}=0$, then  the system will be free from cooperators and defectors ($x=0$ and $z=0$) for all the remaining time $t$. The fraction of punisher $y$ will be
$ \dfrac{(\sigma_{2}-\xi)\big[1+\tanh(c_2+t)\big(\frac{\sigma_{2}}{2}-\frac{\xi}{2}\big)\big]}{2\sigma_{2}-2} \hspace{0.2cm} \text{with} \hspace{0.2cm} \sigma_{2} \ne 1$, where $c_2$ is an initial condition dependent constant. Note that, proceeding to the limit as $t \to \infty$, $y$ will be tending to $\dfrac{\sigma_{2}-\xi}{\sigma_{2}-1} \hspace{0.2cm} \text{with} \hspace{0.2cm} \sigma_{2} \ne 1$ if $\sigma_2 > \xi$. 
	
	\item If $y_{0}=0$ and $z_{0}=0$, then the punishers and defectors ($y=0$ and $z=0$) will die out. The empty intial state with respect to the punishers and defectors actually do not give them opportunity to replicate in future. However, the fraction of cooperator $x$ will be
	$\dfrac{(\sigma_{1}-\xi)\big[1+\tanh(c_3+t)\big(\frac{\sigma_{1}}{2}-\frac{\xi}{2}\big)\big]}{2\sigma_{1}-2} \hspace{0.2cm} \text{with} \hspace{0.2cm} \sigma_{1} \ne 1$,
	where $c_3$ is $x_0$ dependent constant.
	
	\item If only $z_0=0$, then extinction of defectors will happen. Now, if $\sigma_{1}=\sigma_{2}$ and $(1-\sigma_{1})(x+y)+(\sigma_{1}-\xi) \ne 0$, then $y=c_{4}x$, where $c_4$ is a constant. Also, if $\sigma_{1}=\sigma_{2}$ and $(1-\sigma_{1})(x+y)+(\sigma_{1}-\xi) = 0$, then $x=y=0$, and thus all species will die out.
	
	\item If only $y_0=0$, then $y=0$ and generates a punisher-free society. Under this circumstance, $E_5$ will be stable, if
   \[
   \begin{cases}
   \eta_{1} (\beta + \sigma_{3} +2 - \sigma_{1})-\sigma_{3} (1+2\eta_{2})\\-\sigma_1 \eta_{2} - 2\xi < 0, \hspace{0.2cm}  \text{and}\\
   \big[ 2\eta_{1} - \sigma_{1}(\eta_{1}+\eta_{2})-\xi \big] \big[ \beta \eta_{1} - \xi + \sigma_{3} (\eta_{1} -1 -2\eta_{2}) \big]\\
   +\sigma_{1} \eta_{1} (\beta-\sigma_{3})(1+\eta_{2}-\eta_{1})>0.
   \end{cases}
   \]

   \item If only $x_{0} = 0$, then all cooperators will be vanished ($x=0$). Other non-zero components of initial condition $y_0 , z_0\ne 0$ leads to stable stationary point $E_6$, if
	\[
	\begin{cases}	
	\gamma_1 (\beta +\sigma_{3}+2-\sigma_{2})-\sigma_{3}(1+2\gamma_2)\\-\gamma_2(\sigma_2+\delta)-2\xi-\delta<0,\\
	\text{and} \hspace{0.2cm}  \big[2(1-\sigma_{2})\gamma_{1}+(\sigma_2-\xi)\big]\big[(\beta-\sigma_{3}-\delta)\gamma_1\\-2\sigma_{3}(1-\gamma_1+\gamma_2)+(\sigma_{3}-\xi)\big]+(\sigma_{2}+\delta)(1-\gamma_1+\gamma_2)\\\big[2\sigma_{3}(1-\gamma_1+\gamma_2)-(\sigma_{3}-\xi)\big]>0.
	\end{cases}
	\]
	
\end{enumerate}



  \bibliographystyle{elsarticle-num}
  \bibliography{clean_file}





\end{document}